\documentstyle[amssymb,12pt]{article}

\let\large=\normalsize
\def\ft#1#2{{\textstyle{#1\over#2}}}
\def\fft#1#2{{#1\over#2}}
\def\sgn{\,{\rm sgn}}
\def\hoch#1{$\, ^{#1}$}

\begin{document}


\begin{flushright}
CAMS/00-07\\
UM-TH-00-21\\
hep-th/0009212\\
\end{flushright}

\vspace{10pt}

\begin{center}

{\large {\bf Localization of supergravity on the brane}}

\vspace{20pt}

M.J.~Duff\hoch{\dagger1},
James T.~Liu\hoch{\dagger1},
W.A.~Sabra\hoch{\ddagger}

\vspace{10pt}
{\hoch{\dagger} \it
Randall Laboratory, Department of Physics, University of Michigan,\\
Ann Arbor, MI 48109--1120}

\vspace{5pt}

{\hoch{\ddagger} \it
Center for Advanced Mathematical Sciences (CAMS) and Physics Department,\\
American University of Beirut, Lebanon}
\vspace{30pt}

\underline{ABSTRACT}
\end{center}

A supersymmetric Randall-Sundrum brane-world demands that not merely the
graviton but the entire supergravity multiplet be trapped on the brane.
To demonstrate this, we present a complete ansatz for the reduction of 
$(D=5,N=4)$ gauged supergravity to $(D=4,N=2)$ ungauged supergravity in 
the Randall-Sundrum geometry. We verify that it is
consistent to lowest order in fermion terms.  In particular, we show 
how the graviphotons avoid the `no photons on the brane' result because 
they do not originate from Maxwell's equations in $D=5$ but rather 
from odd-dimensional self-duality equations. In the case of the 
fivebrane, the Randall-Sundrum mechanism also provides a new Kaluza-Klein 
way of obtaining chiral supergravity starting from non-chiral.

{\vfill\leftline{}\vfill
\vskip 10pt \footnoterule
{\footnotesize\noindent\hoch{1}
Research supported in part by DOE Grant DE-FG02-95ER40899 Task G.
\vskip  -12pt} \vskip   14pt
}

\newpage


\section{Introduction}

The claim \cite{DLS} that the Type IIB domain wall solution of Bremer
{\it et al.}~\cite{Bremer} provides a supersymmetric realization of the 
Randall-Sundrum \cite{Randall} brane-world naturally requires that not just
the graviton but the whole $(D=4,N=4)$ supergravity multiplet be localized 
on the brane. Indeed, this was implicitly assumed when showing that the 
Maldacena \cite{Maldacena} $(D=4,N=4)$ SCFT%
\footnote{The complete brane theory is 
thus $N=4$ supergravity coupled to $U(\cal N)$ $N=4$ Yang-Mills. 
Since the origin of the Yang-Mills is well understood, however, we focus 
here only on the supergravity.}
gives the same $1/r^{3}$ corrections to the Newtonian potential as those
found by Randall and Sundrum \cite{Duffliu}. Since most treatments of
the brane-world have focussed only on the graviton, however, it would
be desirable to demonstrate the trapping of the entire supermultiplet
explicitly by presenting a complete reduction ansatz of $(D=5,N=8)$ gauged
supergravity to $(D=4,N=4)$ ungauged supergravity in the Randall-Sundrum
geometry. In a first step 
towards this goal, this paper presents the complete reduction ansatz 
for the simpler case of $(D=5,N=4)$ gauged supergravity to $(D=4,N=2)$ 
ungauged supergravity. 

However, in addition to all the no-go theorems 
discussed and refuted in \cite{DLS}, a new one now presents itself%
\footnote{This was first pointed out to us by Eva Silverstein.}.
Unlike gravitons 
obeying an Einstein-Hilbert action, or scalars obeying a Klein-Gordon 
action, photons obeying a Maxwell action in $D=5$ are not localized on the 
brane because the reduction ansatz in the Randall-Sundrum geometry 
leads to a divergent integral \cite{gabad,kaloper}. Yet we certainly require 
the graviphotons in $D=4$ supergravity multiplet to be bound to the 
brane. A second purpose of the present paper, therefore, is to resolve 
this dilemma by invoking some recent results in \cite{lupope}. The no-go
theorem is circumvented because the graviphotons do not originate from a 
Maxwell action in $D=5$ gauged supergravity but rather from an 
``odd-dimensional self-duality'' action \cite{tpvn,grw,ppvn,romans}.  
The $(D=5,N=8)$ gauged supergravity \cite{grw,ppvn} provides 12 two-forms,
transforming as $(6,2)$ of $SO(6)\times SL(2)$.  When paired up in the
reduction of \cite{lupope}, these provide the six graviphotons of the 
$(D=4,N=4)$ theory on the brane. (This $SL(2)$ is then the S-duality 
in $D=4$.)  Similarly, the $(D=5,N=4)$ gauged
supergravity \cite{romans}
has a pair of two-forms, reducing to a single graviphoton of the resulting
$(D=4,N=2)$ theory. Here we focus on the reduction of gauged $(D=5,N=4)$
supergravity, and complete the picture of \cite{lupope} by providing the
reduction ansatz for the fermions and hence demonstrating that the entire
$(D=4,N=2)$ ungauged supergravity multiplet may be confined to the brane.  
These results may be easily generalized to the reduction of the
gauged $(D=5,N=8)$ theory to ungauged $(D=4,N=4)$.

Another novel aspect of localizing supergravity on the brane is that 
of chirality. Just as the $D=10$ $S^{5}$ breathing-mode domain-wall
solution of Bremer {\it et al.}~\cite{Bremer} provides a Type IIB
realization of the Randall-Sundrum threebrane in $AdS_{5}$, so the
$D=11$ $S^{4}$ breathing-mode domain wall solution \cite{Bremer}
provides an M-theory realization of the Randall-Sundrum fivebrane in
$AdS_{7}$.  The resulting theory trapped on the brane is {\it chiral}
$D=6,(N_{+},N_{-})=(2,0)$ supergravity coupled to the SCFT. This thus 
provides a new Kaluza-Klein mechanism, distinct from Horava-Witten 
\cite{Horava:1996qa,Horava:1996ma}, of obtaining a chiral supergravity
theory on the brane starting from a non-chiral theory in the bulk.


\section{Supergravity in the bulk, and reduction of the bosons}
\label{bosons}

In five dimensions, $N=4$ supergravity admits an $SU(2)\times U(1)$ gauging
and an anti-de Sitter vacuum with supergroup $SU(2,2|2)$.  The field content
of this theory consists of a graviton $g_{MN}$, four gravitini
$\Psi_M^a$, $SU(2)\times U(1)$ adjoint gauge fields $A_M^I$ and $a_M$,
two antisymmetric tensors $B_{MN}^\alpha$, four spin-$\fft12$ fields
$\chi^a$, and a real scalar $\phi$.  Indices $a,b$ take values in the
4 of $USp(4)$ and $I,J$ in the adjoint of $SU(2)$, while
$\alpha,\beta=1,2$ correspond to the real vector representation of
$SO(2)\simeq U(1)$.  Indices $M,N,\ldots$ denote five-dimensional
spacetime indices, while $\mu,\nu,\ldots$ denote four-dimensional ones.
This model was constructed in \cite{romans}, and to
leading order in the fermions has a Lagrangian given by:
\begin{eqnarray}
\label{eq:lag}
e^{-1}{\cal {L}} &=&R-\ft{1}{2}(\partial \phi )^{2}
-\ft{1}{4} X^{4}(f_{MN})^{2}-\ft{1}{4}X^{-2}(F_{MN}^{I})^{2}
-\ft{1}{4} X^{-2}(B_{MN}^{\alpha })^{2}
+4g^{2}(X^{2}+2X^{-1})\nonumber\\
&&
+\frac{1}{8g}\epsilon^{MNPQR}\epsilon_{\alpha\beta}
B_{MN}^{\alpha}D_{P}^{\vphantom{\alpha}}B_{QR}^{\beta}
-\frac{1}{8}\epsilon^{MNPQR}
F_{MN}^{I}F_{PQ}^{I}a_{R}^{\vphantom{I}}\nonumber\\
&&
-\ft{1}{2}\bar{\Psi}_M^{a}\gamma^{MNP}D_N\Psi_{P\,a}
+\ft{1}{2}\bar{\chi}^{a}\gamma^MD_M\chi_{a}\nonumber\\
&&
-\ft{3i}{2}\bar{\Psi}_M^{a}\gamma^{MN}T_{ab}\Psi_N^{b}
+ \bar{\Psi}_M^{a}\gamma^MA_{ab}\chi^{b}
+i\bar{\chi}^{a}(\ft{1}{2}T_{ab}-\ft{1}{\sqrt{3}}A_{ab})\chi^{b}\nonumber\\
&&
+\ft{i}{8\sqrt{2}}(H_{MN}^{ab}+\ft{1}{\sqrt{2}}h_{MN}^{ab})
\bar{\Psi}_{a}^P\gamma _{[P}\gamma^{MN}\gamma_{Q]}\Psi_{b}^Q
+\ft{1}{4\sqrt{6}}(H_{MN}^{ab}-\sqrt{2} h_{MN}^{ab})
\bar{\Psi}_{a}^P\gamma ^{MN}\gamma_P\chi_{b}\nonumber\\
&&
+\ft{i}{24\sqrt{2}}(H_{MN}^{ab}-\ft{5}{\sqrt{2}}h_{MN}^{ab})
\bar{\chi}_{a}\gamma^{MN}\chi_{b}
-\ft{i}{2\sqrt{2}}\partial_N\phi\bar{\Psi}_M^{a}\gamma^N\gamma^M\chi_{a},
\end{eqnarray}
where
\begin{eqnarray}
X &=&e^{-\frac{1}{\sqrt{6}}\phi },
\nonumber\\
F_{MN}^{I} &=&2\Bigl(\partial _{[M}^{\vphantom{I}}A_{N]}^{I}
+\fft{g}{\sqrt{2}}\epsilon^{IJK}A_{[M}^{J}A_{N]}^{K}\Bigr),
\nonumber\\
H_{MN}^{ab} &=&X^{-1}\left[F_{MN}^{I}(\Gamma _{I})^{ab}
+B_{MN}^{\alpha }(\Gamma _{\alpha })^{ab}\right],
\nonumber \\
h_{MN}^{ab} &=&X^{2}\Omega ^{ab}f_{MN},
\nonumber\\
T^{ab} &=&\frac{g}{6}(X^{-2}+2X)(\Gamma _{12})^{ab},
\nonumber\\
A^{ab} &=&\frac{g}{\sqrt{3}}(-X^{-2}+X)(\Gamma _{12})^{ab}.
\end{eqnarray}
Here we mostly follow the conventions of \cite{romans} up to a rescaling
of the fields by a factor of $1/2$.  However we use a mostly positive
metric, $\eta_{MN}=(-,+,+,+,+)$.  Additionally, we prefer to label the
$USp(4)$ matrices $(\Gamma_I,\Gamma_\alpha)$ with $\alpha=1,2$ and
$I=3,4,5$ instead.  We have additionally used the freedom
to rescale the $SU(2)\times U(1)$ couplings to set
$g_2=\sqrt{2}g_1=2\sqrt{2}g$.  Note that the bosonic part of (\ref{eq:lag})
agrees with that of \cite{lupope}.

To lowest order, the supersymmetry variations corresponding to the above
Lagrangian are given by
\begin{eqnarray}
\label{eq:evar}
\delta e_M^{r}&=&-\ft{1}{4}\bar{\Psi}_M^{a}\gamma^{r}\epsilon_{a},\\
\label{eq:Avar}
\delta A_M^{I} &=&\Theta_M^{ab}(\Gamma ^{I})_{ab},\\
\label{eq:avar}
\delta a_M &=&\ft{i}{4}X^{-2}(\bar{\Psi}_M^{a}-\ft{2i}{3}
\bar{\chi}^{a}\gamma_M)\epsilon _{a},\\
\label{eq:Bvar}
\delta B_{MN}^{\alpha}&=&2D_{[M}^{\vphantom{ab}}
(\Theta_{N]}^{ab}(\Gamma^{\alpha})_{ab})
-\ft{1}{\sqrt{2}}g\epsilon^{\alpha\beta}(\Gamma_{\beta})_{ab}X^{-1}
(\bar{\Psi}_{[M}^{a}\gamma_{N]}^{\vphantom{a}}\epsilon^{b}
-\ft{i}{2\sqrt{3}}\bar{\chi}^{a}\gamma_{MN}\epsilon^{b}),\\
\label{eq:phivar}
\delta \phi &=&\ft{i}{2\sqrt{2}}\bar{\chi}^{a}\epsilon_{a},\\
\label{eq:psivar}
\delta \Psi_{M\,a}&=&D_M\epsilon_{a}-i\gamma_MT_{ab}\epsilon^{b}
-\ft{i}{12\sqrt{2}}(\gamma_M{}^{NP}-4\delta_M^N\gamma^P)
(H_{NP\,ab}+\ft{1}{\sqrt{2}}h_{NP\,ab})\epsilon^{b},\\
\label{eq:chivar}
\delta \chi_{a} &=&\ft{i}{2\sqrt{2}}\gamma^M\partial_M\phi\epsilon_{a}
+A_{ab}\epsilon^{b}+\ft{1}{4\sqrt{6}}\gamma^{MN}
(H_{MN\,ab}-\sqrt{2}h_{MN\,ab})\epsilon^{b},
\end{eqnarray}
where
\begin{equation}
\Theta_M^{ab}=-\ft{i}{2\sqrt{2}}X(\bar{\Psi}_M^{a}\epsilon^{b}
+\ft{i}{\sqrt{3}}\bar{\chi}^{a}\gamma_M\epsilon^{b}).
\end{equation}

The consistent reduction for the bosonic fields was given in 
\cite{lupope},
where the scalar and one-form gauge fields were set to zero,
\begin{equation}
\label{eq:trunc}
a_M=0,\qquad A_M^{I}=0,\qquad \phi=0,
\end{equation}
while the metric took on a standard Randall-Sundrum form,
\begin{equation}
\label{eq:rsmet}
ds_{5}^{2}=e^{-2k|z|}ds_{4}^{2}+dz^{2},
\end{equation}
where $k$ is always taken positive, corresponding to the `kink-down'
geometry that traps gravity.  As discussed in \cite{bkvp}, supersymmetry
demands the gauge coupling $g$ to flip a sign across the brane, so the
relation between $g$ and $k$ is simply $g=k\sgn(z)$.

The key observation made in \cite{lupope} was that the two-forms
$B_{MN}^\alpha$, which satisfy an odd-dimensional self duality
condition, may be reduced according to
\begin{eqnarray}
B_{\mu\nu}^{\alpha}&=&\ft{1}{\sqrt{2}}e^{-k|z|}
\left\{ F_{\mu\nu},-\star_{4}F_{\mu \nu }\right\},\nonumber\\
B_{\mu z}^\alpha&=&0,
\label{eq:lupope}
\end{eqnarray}
to provide the $D=4$, $N=2$ graviphoton localized on the brane.
Note that this ansatz for the two-form yields $(B_{MN}^\alpha)^2=0$, as
is fitting for a `self-dual' field in general.  For this background, with
the scalar sitting at its fixed point, $X=1$, we find
\begin{equation}
T^{ab}=\frac{g}{2}(\Gamma_{12})^{ab},\qquad A^{ab}=0,
\end{equation}
while many other terms drop out because of the trivial one-form potentials.
Additionally, we have
\begin{equation}
H_{\mu\nu}^{ab}=B_{\mu \nu }^{\alpha }(\Gamma _{\alpha })^{ab},\qquad
h_{\mu\nu}^{ab}=0.
\end{equation}
It was shown in \cite{lupope} that the bosonic equations of motion reduce to
\begin{eqnarray}
\label{eq:beom}
R_{\mu \nu }^{(4)} &=&\ft{1}{2}{}(F_{\mu \lambda }F_{\nu}{}^{\lambda}
-\ft{1}{4}g_{\mu \nu }F^{2}),\nonumber\\
dF &=&0,\qquad d\star_4F_{\mu \nu }=0.
\end{eqnarray}
This is the expected form of the equations corresponding to the graviton
and graviphoton of ungauged $D=4$, $N=2$ supergravity.


\section{The reduction ansatz for the fermions}

We now turn to consideration of the fermion fields.  With the expectation
that pure $N=2$ supergravity in four dimensions does not contain any
spin-$\fft12$ fields, it is natural to set
\begin{equation}
\label{eq:chians}
\chi_a=0.
\end{equation}
What remains to be determined is the ansatz for the gravitino $\Psi_M^a$
as well as the form of the supersymmetry parameter $\epsilon$.  We start
with the spin-$\fft12$ variation, (\ref{eq:chivar}), which on this background
takes the form
\begin{eqnarray}
\delta \chi_{a}&=&\ft{1}{4\sqrt{6}}\gamma^{MN}B_{MN}^{\alpha}
(\Gamma_{\alpha})_{ab}\epsilon^{b} \nonumber\\
&=&\ft{1}{8\sqrt{3}}e^{k\mid z\mid }\hat{\gamma}^{\mu \nu }
(F_{\mu\nu}\Gamma_{1}-\star_{4}F_{\mu\nu}\Gamma_{2})_{ab}\epsilon^{b},
\end{eqnarray}
where four-dimensional Dirac matrices, denoted with a caret, have curved
space indices corresponding to the four-dimensional vierbein.  A standard
manipulation using the identity
\begin{equation}
\ft{1}{2}\epsilon _{\mu\nu\lambda\sigma}\hat\gamma^{\lambda\sigma}
=i\hat\gamma_{\mu\nu}\gamma^{5}
\end{equation}
then brings the spin-$\fft12$ variation to the final form
\begin{equation}
\delta\chi_{a}=\ft{1}{8\sqrt{3}}e^{k|z|}\hat{\gamma}^{\mu\nu}F_{\mu\nu}
(\Gamma_{1})_{ab}(1-i\gamma^{5}\Gamma^{12})^b{}_c\epsilon^{c}.
\end{equation}
We keep in mind that $\gamma^5$ is the four-dimensional chirality
operator, $\gamma^5\equiv i\gamma^0\gamma^1\gamma^2\gamma^3$.

Since in the reduction we have set $\chi_{a}=0$ in (\ref{eq:chians}),
the above transformation is consistent with the ansatz only if it can be set
to zero.  This implies that the surviving supersymmetry on the brane
must satisfy
\begin{equation}
(1-i\gamma^{5}\Gamma^{12})\epsilon=0.
\end{equation}
This result is consistent with the BPS nature of the Randall-Sundrum
brane, where the `kink' breaks half of the bulk supersymmetries.

Proceeding with the gravitino transformation, (\ref{eq:psivar}), we find
\begin{equation}
\delta\Psi_{M\,a}=\nabla_M\epsilon_{a}-\ft{i}{2}g\gamma_M
(\Gamma_{12})_{ab}\epsilon^{b}-\ft{i}{12\sqrt{2}}
(\gamma_M{}^{NP}-4\delta_M^N\gamma^P)
(B_{NP}^{\alpha})(\Gamma_{\alpha})_{ab}\epsilon^{b}.
\end{equation}
The $z$ component is then
\begin{equation}
\delta\Psi_{z\,a}=\partial_{z}\epsilon_{a}-\ft{i}{2}g\gamma_{z}
(\Gamma_{12})_{ab}\epsilon^{b}-\ft{i}{24}e^{k|z|}\gamma_{z}
\hat{\gamma}^{\mu\nu}F_{\mu\nu}(\Gamma_{1})_{ab}
(1-i\gamma^{5}\Gamma^{12})^b{}_c\epsilon ^{c},
\end{equation}
which may be rewritten in the form
\begin{equation}
\delta\Psi_{z}=\left[e^{-\fft12gz}\partial_{z}e^{\fft12gz}
-\ft12\left(g+\ft{i}{12}e^{k|z|}\gamma_{z}\hat{\gamma}^{\mu\nu}
F_{\mu\nu}\Gamma_{1}\right)(1-i\gamma^{5}\Gamma^{12})\right]\epsilon,
\end{equation}
where all symplectic indices have been suppressed.  We have furthermore
made the identification $\gamma_{z}=-\gamma^{5}$, corresponding to the
$D=5$ relation $\gamma^0\gamma^1\gamma^2\gamma^3\gamma^{\overline{z}}=i$.
Since $g=k\sgn(z)$, it is now clear that the above transformation
will vanish for a supersymmetry parameter of the form
\begin{equation}
\label{eq:epsred}
\epsilon(x,z)=\ft12e^{-\fft12k|z|}(1+i\gamma ^{5}\Gamma^{12})
\hat\epsilon(x).
\end{equation}
The spinors $\hat\epsilon(x)$ correspond to the surviving supersymmetry on
the brane.  The exponential factor $e^{-\fft12k|z|}$ agrees with that
anticipated in \cite{lupope}.

We now reduce the four-dimensional component of the gravitino transformation
to obtain
\begin{eqnarray}
\delta\Psi_{\mu}&=&\Bigl[\hat\nabla_\mu
-\ft{i}8(\hat\gamma_\mu{}^{\nu\rho}-2\delta_\mu^\nu\hat\gamma^\rho)
F_{\nu\rho}\Gamma_1\nonumber\\
&&\qquad-\ft12\left(ge^{-k|z|}\hat\gamma_\mu\gamma^z
-\ft{i}{6}(\hat\gamma_\mu{}^{\nu\rho}-\delta_\mu^\nu\hat\gamma^\rho)
F_{\nu\rho}\Gamma_1\right)(1-i\gamma^5\Gamma^{12})\Bigr]\epsilon.
\end{eqnarray}
This indicates that the proper reduction of the gravitino is given by
\begin{eqnarray}
\label{eq:grared}
\Psi_\mu(x,z)&=&\ft12e^{-\fft12k|z|}(1+i\gamma ^{5}\Gamma^{12})
\hat\psi_\mu(x),\nonumber\\
\Psi_z(x,z)&=&0,
\end{eqnarray}
so that
\begin{equation}
\delta\hat\psi_{\mu}=\left[\hat\nabla_{\mu}
-\ft{i}8(\hat\gamma_\mu{}^{\nu\rho}-2\delta_\mu^\nu\hat\gamma^\rho)
F_{\nu\rho}\Gamma_1\right]\hat\epsilon.
\end{equation}
%


\section{Consistency of the fermion ansatz}

It was shown in \cite{lupope} that the consistent reduction of the
bosonic sector is compatible with the field truncation (\ref{eq:trunc}).
In the presence of fermions, it is also necessary to verify that the
supersymmetry variations of the bosons respect this truncation.
Using the above reduction ansatz and the Majorana transpose
\begin{equation}
\overline\epsilon = \ft12e^{-\fft12k|z|}\overline{\hat\epsilon}
(1-i\gamma^5\Gamma^{12}),
\end{equation}
one can easily check from (\ref{eq:Avar}), (\ref{eq:avar}) and
(\ref{eq:phivar}) that
\begin{eqnarray}
\delta A_{\mu}^{I} &=&0,\qquad \delta A_z^I=0, \nonumber\\
\delta a_{\mu} &=&0,\qquad \delta a_z=0, \nonumber\\
\delta \phi &=&0.
\end{eqnarray}
Furthermore, from (\ref{eq:evar}) and (\ref{eq:Bvar}), we find that the
$z$ component transformations vanish,
\begin{eqnarray}
\delta e_{z}^{\bar{z}} &=&0, \nonumber\\
\delta B_{\mu z}^\alpha &=&0,
\end{eqnarray}
while the four-dimensional components reduce to give the surviving
supersymmetry transformations for the bosonic fields
\begin{eqnarray}
\delta e_\mu^r&=&-\ft{1}{4}\bar{\hat\psi}_{\mu}\gamma^r\hat\epsilon,
\nonumber\\
\delta A_\mu&=&-\ft{i}{2}\bar{\hat\psi}_{\mu}\Gamma^1\hat\epsilon.
\end{eqnarray}
In fact, the reduction of $\delta B^\alpha_{\mu\nu}$ leads to a
variation on the field strength, $\delta F_{\mu\nu}$.  The
transformation on the potential, given above, is deduced (up to a gauge
transformation) by removing a derivative on both sides of the actual
expression.

Finally, we examine the consistency and reduction of the fermion equations
of motion.  For the spin-$\fft12$ equation, we have
\begin{eqnarray}
\gamma^M D_M\chi_a &=&\gamma^M A_{ab}\Psi_M^b
+2i(\ft12T_{ab}-\ft1{\sqrt{3}}A_{ab})\chi^b
+\ft{i}{4\sqrt{6}}(H_{MN}^{ab}-\sqrt{2}h_{MN}^{ab})\gamma_P
\gamma^{MN}\Psi_b^P\nonumber\\
&&+\ft{i}{12\sqrt{2}}(H_{MN}^{ab}-\ft5{\sqrt{2}}h_{MN}^{ab})
\gamma^{MN}\chi_b
-\ft{i}{2\sqrt{2}}\partial_N\phi\gamma^M\gamma^N\Psi_M^a,
\end{eqnarray}
which, for the reduction ansatz, simplifies to the condition
\begin{equation}
\gamma_\rho\gamma^{\mu\nu}H_{\mu\nu}^{ab}\Psi_b^\rho=0.
\end{equation}
This is satisfied since the combination $\gamma^{\mu\nu}H_{\mu\nu}^{ab}$
is projected out on the trapped gravitino given by (\ref{eq:grared}).

On the other hand, the five-dimensional gravitino equation of motion is
expected to reduce to its four-dimensional counterpart.  From
(\ref{eq:lag}), we find
\begin{eqnarray}
\gamma^{MNP}D_N\Psi_{P\,a}&=&
-3i\gamma^{MN}T_{ab}\Psi_N^b+\gamma^M A_{ab}\chi^b
+\ft{i}{4\sqrt{2}}(H^{PQ\,ab}+\ft1{\sqrt{2}}h^{PQ\,ab})
\gamma^{[M}\gamma_{PQ}\gamma^{N]}\Psi_{N\,b}\nonumber\\
&&+\ft{1}{4\sqrt{6}}(H_{PQ}^{ab}-\sqrt{2}h_{PQ}^{ab})
\gamma^{PQ}\gamma^M\chi_b
-\ft{i}{2\sqrt{2}}\partial_N\phi\gamma^N\gamma^M\chi_a.
\end{eqnarray}
Using the reduction ansatz, this becomes
\begin{equation}
\label{eq:greom}
\gamma^{MNP}\nabla_N\Psi_P=-\ft{3i}{2}g\gamma^{MN}
\Gamma^{12}\Psi_N+\ft{i}{4\sqrt{2}}B^{\rho\lambda\,\alpha}\gamma^{[M}
\gamma_{\rho\lambda}\gamma^{N]}\Gamma_\alpha\Psi_N.
\end{equation}
The five-dimensional covariant derivative $\nabla_M$ picks up a
contribution from the warped metric background, (\ref{eq:rsmet}), which
cancels against the first term on the right hand side of (\ref{eq:greom}).
Picking $\mu$ to be a four-dimensional index, the resulting gravitino
equation of motion reduces to
\begin{equation}
\label{eq:4geom}
\hat\gamma^{\mu\nu\rho}\hat\nabla_\nu\hat\psi_\rho
=\ft{i}4F^{\rho\lambda}(\hat\gamma^{[\mu}\hat\gamma_{\rho\lambda}
\hat\gamma^{\nu]})\Gamma_1\hat\psi_\nu.
\end{equation}
For the $z$ component of (\ref{eq:greom}), a similar computation
indicates
\begin{equation}
\hat\gamma^{\mu\nu}\hat\nabla_\mu\hat\psi_\nu=
\ft{i}8F_{\rho\lambda}
\hat\gamma^{\rho\lambda}\hat\gamma^\mu\Gamma_1\hat\psi_\mu,
\end{equation}
which is equivalent to the $\hat\gamma_\mu$ contraction of
(\ref{eq:4geom}).


\section{Evading the `no photons on the brane' result}

Let us first recall the arguments of \cite{gabad,kaloper} showing that 
photons originating from a Maxwell action in $D=5$ are not trapped on 
the brane. Starting from 
\begin{equation}
S_{\rm Maxwell}¥=\int d^{4}x\,dz\sqrt{-g_{(5)}}\left[
-\fft{1}{4}g^{MP}¥g^{NQ}¥F_{MN}¥F_{PQ}¥\right],
\end{equation}
and making the reduction ansatz (\ref{eq:rsmet}) for the metric and
\begin{equation}
A_{M}(x,z)=\{A_{\mu}(x), 0\}
\label{photonansatz}
\end{equation}
for the photon, we find
\begin{equation}
S_{\rm photon}¥=\int d^{4}x¥\sqrt{-g}\left[
-\fft{1}{4}g^{\mu\rho}¥g^{\nu\sigma}¥F_{\mu\nu}¥F_{\rho\sigma}¥\right]
\int_{-\infty}^{\infty}¥dz,
\end{equation}
for which the $z$ integral diverges. This is due to the two powers of the 
inverse metric appearing in the Maxwell action and contrasts with both 
the Einstein-Hilbert and Klein-Gordon actions
\begin{equation}
S_{\rm Einstein-Hilbert-Klein-Gordon}
=\int d^{4}x\,dz\sqrt{-g_{(5)}}\left[g^{MN}R_{MN}
-\fft{1}{2}g^{MN}¥\partial_{M}¥\phi\partial_{N}¥\phi\right],
\end{equation}
which have only one power of the inverse metric and hence yield
\begin{equation}
S_{\rm gravity-scalar}¥=\int d^4{x}¥\sqrt{-g}\left[g^{\mu\nu}¥R_{\mu\nu}¥-
\frac{1}{2}g^{\mu\nu}¥\partial_{\mu}¥\phi\partial_{\nu}¥\phi \right]
\int_{-\infty}^{\infty}dz\, e^{-2k|z|},
\label{eq:gsconv}
\end{equation}
for which the $z$ integral converges. This result was interpreted in
\cite{kaloper} as a charge-screening effect. However, supersymmetry 
rules out any such effect for the graviphotons in the $N=4$ 
supergravity multiplet, whose action must lead to the same convergent 
integral as the graviton and graviscalars.  The ansatz
(\ref{photonansatz}) in any case is ruled out in the Randall-Sundrum
geometry (\ref{eq:rsmet}) by the bulk Einstein equation \cite{hodge}.
Furthermore, attempts to modify the ansatz, {\it e.g.}~through inclusion
of a converging $z$-dependent factor
\begin{equation}
A_{M}(x,z)=\{ e^{-ak|z|}¥A_{\mu}(x), 0 \},
\end{equation}
will not cure the problem because this would not even solve the bulk
Maxwell equations \cite{hodge}.

The resolution, as we have seen in section (\ref{bosons}), is to 
start instead with the 2-form field $B_{MN}^\alpha$, and then make
the {\it $z$-dependent} Lu-Pope \cite{lupope} ansatz (\ref{eq:lupope}) to
arrive at graviphotons on the brane.  Note, however, that the argument
given above for reduction of the action no longer directly applies in
this case, as the reduction ansatz (\ref{eq:lupope}) is on the field
strength $F_{\mu\nu}$ and not on the potential.  As a result, a
straightforward reduction of the the odd-dimensional self-duality action
for the 2-form
\begin{equation}
\label{eq:odsde}
S_{\rm 2-form}=\int 
d^{4}x\,dz\sqrt{-g_{(5)}}\left[-\frac{1}{4}g^{MP}g^{NQ}¥B^{\alpha}_{MN}
B^\alpha_{PQ}
+\frac{1}{8g}\epsilon^{MNPQR}\epsilon_{\alpha\beta}B^{\alpha}_{MN}D_{P}
B^{\beta}_{QR}¥\right],
\end{equation}
would then yield a vanishing four-dimensional action because of the
`self-duality'.  That this is expected may be seen by considering, for
example, the $(D=5, N=8)$ case, whereupon the $SL(2)$
symmetry of the five-dimensional action reduces to the $SL(2)$
S-duality of the $(D=5,N=4)$ theory on the brane.  This S-duality,
however, does not appear in the Lagrangian, but only at the level of
the equations of motion.  Thus the reduced Lagrangian has no choice but
to vanish.

Nevertheless, the essence of the argument may be seen by relaxing the
ansatz, (\ref{eq:lupope}), to
\begin{equation}
\label{eq:modlp}
B_{\mu\nu}^{\alpha}=\ft{1}{\sqrt{2}}e^{-k|z|}F_{\mu\nu}^{(\alpha)},
\qquad B_{\mu z}^\alpha=0.
\end{equation}
The duality condition, $F_{\mu\nu}^{(1)} = *_4 F_{\mu\nu}^{(2)}$, then
arises from the odd-dimensional self-duality equation following from
(\ref{eq:odsde}).  Inserting (\ref{eq:modlp}) into the 2-form action
yields
\begin{equation}
S_{\rm graviphoton}= \int d^{4}x¥\sqrt{-g}\left[
-\frac{1}{8}g^{\mu\rho}¥g^{\nu\sigma}¥F^{(\alpha)}_{\mu\nu}
F^{(\alpha)}_{\rho\sigma}¥\right]
\int_{-\infty}^{\infty}dz\, e^{-2k|z|},
\end{equation}
for which the $z$ integral has the same convergence as (\ref{eq:gsconv}).
Although there are 
two powers of the inverse metric, this is compensated by the 
$z$-dependence of the ansatz.  This action of course vanishes upon the
substitution $F_{\mu\nu}^{(1)}=*_4F_{\mu\nu}^{(2)}$, since $(*_4)^2=-1$.

\section{Discussion}

As anticipated in \cite{lupope}, we have demonstrated that the bosonic
reduction ansatz in \cite{lupope} may be straightforwardly extended to
include the fermion content as well.
Thus we have shown that the $D=5$, $N=4$ gauged $SU(2)\times U(1)$
supergravity has a consistent reduction to pure ungauged $N=2$
(Einstein-Maxwell) supergravity in four dimensions.  This reduction
corresponds to the localization of a complete supergravity multiplet on
the Randall-Sundrum brane, thus confirming the supersymmetry of the
scenario.

Collecting the above results, the reduction is summarized by setting
\begin{eqnarray}
ds_5^2&=&e^{-2k|z|}ds_4^2+dz^2,
\nonumber\\
B_{\mu\nu}^{\alpha}&=&\ft{1}{\sqrt{2}}e^{-k|z|}
\left\{ F_{\mu\nu},-\star_{4}F_{\mu \nu }\right\},
\nonumber\\
\Psi_\mu&=&\ft12e^{-\fft12k|z|}(1+i\gamma ^{5}\Gamma^{12})\hat\psi_\mu,
\end{eqnarray}
with all other fields set to zero.  The reduced equations of motion,
(\ref{eq:beom}) and (\ref{eq:4geom}), may be derived from the following
four-dimensional Lagrangian:
\begin{equation}
e^{-1}{\cal L}=R-\ft14F_{\mu\nu}^2-\ft{1}2\overline{\hat\psi}_\mu
\hat\gamma^{\mu\nu\rho}\hat\nabla_\nu\hat\psi_\rho
+\ft{i}8\overline{\hat\psi}_\mu(\hat\gamma^{[\mu}F\cdot\hat\gamma
\hat\gamma^{\nu]})\Gamma_1\hat\psi_\nu,
\end{equation}
with corresponding supersymmetry transformations
\begin{eqnarray}
\delta e_\mu^r&=&-\ft{1}{4}\bar{\hat\psi}_{\mu}\gamma^r\hat\epsilon,
\nonumber\\
\delta A_\mu&=&-\ft{i}{2}\bar{\hat\psi}_{\mu}\Gamma^1\hat\epsilon,
\nonumber\\
\delta\hat\psi_{\mu}&=&\left[\hat\nabla_{\mu}
-\ft{i}8(\hat\gamma_\mu{}^{\nu\rho}-2\delta_\mu^\nu\hat\gamma^\rho)
F_{\nu\rho}\Gamma_1\right]\hat\epsilon.
\end{eqnarray}
The four-dimensional supersymmetry transformation parameter
$\hat\epsilon$ is related to the five-dimensional one by
\begin{equation}
\epsilon=\ft12e^{-\fft12k|z|}(1+i\gamma^{5}\Gamma^{12})\hat\epsilon.
\end{equation}
Up to a conversion between five-dimensional symplectic-Majorana spinors
and their four-dimensional (Majorana) counterparts, this gives precisely the
$D=4$, $N=2$ theory.

Note that, while the absolute value `kink' at $z=0$ is necessary for the
trapping of gravity by the brane, we have mostly ignored its effects.  In
particular, we have not considered the source terms that must necessary be
present in the Lagrangian.  Presumably such terms may be interpreted in the
original type IIB theory as appropriate (supersymmetric) D3-brane source
couplings.  A proper understanding of the sources would require such a
higher dimensional point of view, as indicated in \cite{DLS,bkvp}.

We have seen how a possible no-go theorem for photons on the brane is 
circumvented for the graviphotons because they originate from odd-dimensional 
self-duality equations rather than Maxwell equations. Consistently 
with this, the ansatz for the massless modes requires that we set to 
zero those photons $a_{\mu}$ and $A_{\mu}^{I}$ which do obey Maxwell's 
equations and also those Kaluza-Klein photons arising from the $g_{\mu z}$
components of the metric. Of course, there will be massive modes 
arising from these and all the other fields that will fall into a continuum
of short $(D=4,N=4)$ massive supermultiplets.
 
Although the reduction of $N=2$ gauged supergravity gives rise to an
$N=1$ theory without any graviphotons, the trapped graviton still has a
gravitino as a superpartner.  This theory can in fact be obtained from a
truncation of the $N=4$ Lagrangian, (\ref{eq:lag}), with the $N=2$
graviphoton given by setting $a_\mu = A_\mu^3/\sqrt{2}$.  The reduction to
four-dimensions then follows, with only (\ref{eq:rsmet}), (\ref{eq:epsred})
and (\ref{eq:grared}) active, while setting all other fields to zero.
Note that the $N=1$ gravitino is Majorana, so this reduction to pure
supergravity does not generate chirality.  The situation is different,
however, in reducing maximal gauged supergravity in seven dimensions
\cite{ppvn7,ppvnw7}, as this would give rise to a six-dimensional $(2,0)$
theory on the brane.  Like in the
Ho\v{r}ava-Witten model \cite{Horava:1996qa,Horava:1996ma}, this generates
chirality on the brane from a non-chiral theory in the bulk.
We recently became aware of \cite{clp}, which overlaps with the present
results.

\section*{Acknowledgments}
JTL and WAS wish to thank the hospitality of Khuri Lab at The Rockefeller
University where much of the work was performed. MJD acknowledges 
conversations with Lisa Randall, Eva Silverstein, and Edward Witten.

\end{document}